\documentclass[aps,prl,twocolumn,superscriptaddress,nofootinbib,floatfix]{revtex4-2}
\usepackage[T1,OT1]{fontenc}
\DeclareTextCommandDefault{\k}[1]{{\fontencoding{T1}\selectfont\k{#1}}}
\usepackage{amsmath,amssymb,dsfont,graphicx,xcolor,hyperref}
\hypersetup{hidelinks}
\graphicspath{{figs/}}
\usepackage{comment}

\newcommand{\Yobs}{Y_B^{\rm obs}}

\begin{document}

\title{
Sterile Neutrino Dark Matter Cries for GeV Heavy Neutral Leptons}

\author{Marco Drewes}
\affiliation{Centre for Cosmology, Particle Physics and Phenomenology (CP3), Universit\'e catholique de Louvain, B-1348 Louvain-la-Neuve, Belgium}
\affiliation{Physik-Department, Technische Universit\"at M\"unchen, D-85748 Garching, Germany}
\author{Yannis Georis}
\affiliation{Kavli IPMU (WPI), UTIAS, University of Tokyo, Kashiwa, Chiba 277-8583, Japan}
\affiliation{Center for Data-Driven Discovery, Kavli IPMU (WPI), UTIAS,
The University of Tokyo, Kashiwa, Chiba 277-8583, Japan}
\affiliation{Centre for Cosmology, Particle Physics and Phenomenology (CP3), Universit\'e catholique de Louvain, B-1348 Louvain-la-Neuve, Belgium}
\author{Juraj Klari\'c}
\affiliation{University of Zagreb, Faculty of Science, Department of Physics, 10000 Zagreb, Croatia}
\author{Mikhail Shaposhnikov}
\affiliation{Institute of Physics, \'Ecole Polytechnique F\'ed\'erale de Lausanne (EPFL), CH-1015 Lausanne, Switzerland}
\author{Inar Timiryasov}
\affiliation{Niels Bohr Institute, University of Copenhagen, Jagtvej 155A, DK-2200 Copenhagen, Denmark}

\date{\today}

\begin{abstract}
The ordinary and dark matter in the Universe may share a common origin in the framework of a minimal extension of the Standard Model by three right-handed neutrinos. A pair of such heavy neutral leptons (HNLs) can give masses to neutrinos, generate the baryon asymmetry at the electroweak scale and produce a large lepton asymmetry at the QCD scale, which is then resonantly transformed into an abundance of the third state that acts as sterile neutrino dark matter. The minimality of this setup, known as Neutrino Minimal Standard Model ($\nu$MSM), makes it highly predictive. Earlier attempts to pin down the properties of the HNLs based on the requirement to simultaneously explain the neutrino masses as well as the baryonic and dark matter abundances were hampered by an incomplete understanding of the quantum kinetic equations describing the HNLs throughout cosmic history. We perform the first parameter scan of the $\nu$MSM in which this shortcoming has been overcome. We find that the $\nu$MSM provides a common explanation for all aforementioned phenomena only within a well-defined parameter space that is limited in all directions,  with HNL masses  kinematically accessible by the SHiP experiment as well as searches at the LHC and future lepton colliders. Further predictions for the HNL decay branching ratios, CP-violation, and lepton number violation (LNV) make the model highly testable, bringing a discovery of the common origin of neutrino masses and all matter in the Universe within reach of existing and planned experiments. 
\end{abstract}

\maketitle
\emph{Introduction}--The matter-antimatter asymmetry of the Universe \cite{Canetti:2012zc} and the identity of its dark matter \cite{Cirelli:2024ssz} are usually treated as separate problems. The $\nu$MSM \cite{Asaka:2005an, Asaka:2005pn}, 
an extension of the Standard Model of particle physics by three right-handed neutrinos,
relates both of them to the origin of neutrino masses.  
The lightest heavy neutrino, $N_1$, with mass $M_1$ at the keV scale and  mixing angles $\theta_{\alpha 1}$ with the active neutrinos of flavour $\alpha = e, \mu, \tau$, 
is the dark-matter (DM) candidate.  The heavier two, $N_2$ and $N_3$, form a pair with nearly-degenerate masses $M_{2,3} = M \pm \Delta M_{\rm phys}/2$, with a total active--sterile mixing $U^2=\sum_{\alpha i}|\theta_{\alpha i}|^2$; 
their CP-violating interactions 
 Yukawa interactions $F_{\alpha i}$
 to the Higgs field and SM leptons $\ell_{L \alpha}$ 
generate the baryon asymmetry of the universe (BAU) \cite{Akhmedov:1998qx, Asaka:2005pn} and  account for the observed neutrino masses \cite{Minkowski:1977sc, GellMann:1980vs, Mohapatra:1979ia, Yanagida:1980xy, Schechter:1980gr, Schechter:1981cv}.  With no other new physics added, the $\nu$MSM provides a self-consistent description of Nature up to the Planck scale \cite{Bezrukov:2012sa} in which the Higgs boson can drive cosmic inflation, leaving the Universe radiation-dominated from temperatures $T$ well above the electroweak scale, with zero initial abundances of all HNLs \cite{Bezrukov:2008ut}.

Economy has a price: The same pair $N_{2,3}$ must yield the correct pattern of neutrino masses and mixings, the BAU, and the conditions for DM production.
The latter is the hardest part.  Non-resonant active--sterile conversion \cite{Dodelson:1993je, Asaka:2006nq} cannot supply enough $N_1$, as X-ray limits on radiative $N_1$ decay together with structure-formation bounds on its free-streaming close that window \cite{Boyarsky:2018tvu, Drewes:2016upu, Venumadhav:2015pla, Schneider:2016uqi}.  
Resonant conversion of lepton chemical potentials $\mu_\alpha$ into $N_1$ \cite{Shi:1998km, Laine:2008pg} is feasible, 
but requires a lepton asymmetry about six orders of magnitude larger than the baryon asymmetry \cite{Venumadhav:2015pla,Ghiglieri:2015jua}, present around the QCD epoch at $T \sim 100$ MeV. That asymmetry must be created at  $T < T_{\rm sph} \simeq132\,$GeV \cite{DOnofrio:2014rug, Eijima:2017cxr}, otherwise SM sphaleron reactions \cite{Kuzmin:1985mm} would tie it to the baryon asymmetry.

Pinning down the $\nu$MSM parameter space that delivers all of this is very challenging, as it requires a precise description of the non-equilibrium dynamics of HNLs throughout cosmic history -- their creation from the plasma of SM particles, coherent oscillations, the loss of quantum coherence, flavour and chirality effects, and non-linear back-reactions on the produced lepton asymmetries and DM abundance.
Refs.~\cite{Ghiglieri:2019kbw, Gorbunov:2025nqs, Akita:2025txo, Vogel:2025aut}  accounted for many effects in DM production, but assumed {\em preexisting} lepton asymmetries $\mu_\alpha$ at the QCD scale, leaving the connection between $N_1$ and the heavier pair $N_{2,3}$ open.
While the general form of the quantum kinetic equations (QKEs) required to compute the $\mu_\alpha$ has been known for a long time \cite{Asaka:2005pn} and a first scan of the $\nu$MSM parameter space was performed over a decade ago \cite{Canetti:2012kh, Canetti:2012vf}, more recent works have not only revised the computation of the relevant interaction rates \cite{Biondini:2017rpb,Laine:2022pgk}, but also revealed a number of qualitatively new aspects that were missed in earlier works, including the  
importance of chirality-flip effects \cite{Eijima:2017anv, Ghiglieri:2017gjz} and the feedback of the $\mu_\alpha$ on the $N_{2,3}$ evolution \cite{Ghiglieri:2020ulj}. To date, no  parameter space exploration 
taking into account these effects
has been performed 
to identify the parameter regime where right-handed neutrinos alone can simultaneously explain the neutrino masses along with the origin of dark and ordinary matter.

This Letter closes the gap. The results are unexpected and exciting. 
Using a DM mass of $M_1=7$~keV~\cite{Bulbul:2014sua, Boyarsky:2014jta, Boyarsky:2014ska} as a benchmark,  we found that the requirement to create the $\mu_\alpha$ needed for DM production narrows the mass to ${M < 50~ {\rm GeV}}$.
In the mass range accessible in meson decays the mixing angles $U^2$ are at least an order of magnitude above the seesaw floor. This is great news for future experiments designed to search for HNLs, such as SHiP \cite{Alekhin:2015byh,SHiP:2025ows} and FCC-ee \cite{FCC:2025lpp} or CEPC \cite{Ai:2025cpj} at the Z-pole \cite{Blondel:2014bra}. Theoretically, this pattern indicates the existence of a global lepton-number symmetry proposed in \cite{Shaposhnikov:2006nn}.

\emph{Formulation of the problem}--
Lepton asymmetries $\mu_\alpha$ are created by $N_{2,3}$ whenever their abundances deviate from equilibrium \cite{Sakharov:1967dj}, which subsequently happens during their approach to equilibrium, freeze-out and decay \cite{Klaric:2020phc,Klaric:2021cpi}.  
If the $\mu_\alpha$ generated during HNL decays source the baryon asymmetry \cite{Fukugita:1986hr}, experimentally accessible $M$ require a near-degenerate HNL mass spectrum  \cite{Pilaftsis:2003gt}.
While baryogenesis during the approach to equilibrium in principle does not rely on a mass degeneracy \cite{Drewes:2012ma,Abada:2018oly,Drewes:2021nqr}, in the $\nu$MSM  the requirement $\Delta M_{\rm phys} \ll M$ is imposed by the necessity to generate $\mu_\alpha$ greatly exceeding the baryon asymmetry at the QCD scale \cite{Shaposhnikov:2008pf,Canetti:2012vf,Canetti:2012kh}, needed to create the conditions for resonant sterile neutrino SM production \cite{Venumadhav:2015pla,Ghiglieri:2015jua,Ghiglieri:2020ulj}.
Baryogenesis in the $\nu$MSM is now well understood \cite{Akhmedov:1998qx, Asaka:2005pn, Shaposhnikov:2008pf, Canetti:2012kh, Canetti:2012vf, Drewes:2016lqo, Hernandez:2016kel, Drewes:2016gmt, Eijima:2017anv, Ghiglieri:2017gjz, Eijima:2018qke, Ghiglieri:2018wbs, Bodeker:2019rvr, Klaric:2020phc, Klaric:2021cpi,Hernandez:2022ivz} and leads to testable predictions for the HNL properties \cite{Shaposhnikov:2006nn, Gorbunov:2007ak, Shaposhnikov:2008pf,  Drewes:2016gmt, Hernandez:2016kel,Antusch:2017pkq,Hernandez:2022ivz}.  
In the present work we hence focus on the generation of large $\mu_\alpha$ at $T<T_{\rm sph}$ within the viable baryogenesis parameter region. 

Qualitatively, generating large lepton asymmetries requires $\omega_{\rm osc} \simeq \Gamma_N$ and  $\Gamma_N\simeq H$ \cite{Shaposhnikov:2008pf, Canetti:2012vf, Canetti:2012kh}, where $\omega_{\rm osc}$ is the frequency of $N_{2,3}$ oscillations, $\Gamma_N$ is the rate of HNL interactions, and $H$ is the Hubble rate.
The first equality ensures resonance in $N_{2,3}$ oscillations, amplifying the CP-violation (familiar from $K^0-\bar{K}^0$ oscillations and proposed for baryogenesis in \cite{Kuzmin:1970nx});
the second equality maximises the departure from thermal equilibrium. 
The large asymmetry must be present at the right time, $T\gtrsim 100$ MeV, when the conversion into sterile neutrino DM occurs. These conditions shrink the parameter space enormously: for the decay of non-relativistic HNLs at $T\sim 100$ MeV, the physical mass splitting between $N_{2,3}$ in vacuum should be as small as $10^{-6}$ eV to be in the resonance \cite{Roy:2010xq}.

To identify the viable parameter space for baryogenesis and DM production we solve the QKEs summarised in supplementary material
for the variables that change much slower than the rates of the perturbative SM interactions: the
density matrices $\rho_N$ and $\bar{\rho}_N$ of two $N_{2,3}$ helicity states, 
the occupancies of the two helicity states of $N_1$, 
three flavoured  asymmetries $Y_{\Delta_\alpha}$ that are tied to the lepton chemical potentials $\mu_\alpha$ via a susceptibility matrix and include spectator effects \cite{Bodeker:2015zda}, 
and the baryon yield $Y_B$. 
An extra outcome of this program are the spectra of the produced DM sterile neutrinos. 
to be confronted with more detailed but less robust cosmological structure-formation probes such as Lyman-$\alpha$ \cite{Baur:2017stq, Garzilli:2019qki} and satellite-count bounds \cite{Nadler:2021dft, Dekker:2021scf, Irsic:2023equ}; we leave this study for the future.

\emph{The mechanism for late lepton asymmetry generation}--
We find that there are two distinct mechanisms to produce sizeable $\mu_\alpha$ at $T\lesssim 1$ GeV, 
to which we refer to as \emph{resonant decay mechanism} and 
the \emph{ignition mechanism}. 
The \emph{resonant decay mechanism} is essentially a low-temperature version of resonant leptogenesis \cite{Pilaftsis:2003gt} operative at $\omega_{\rm osc} \simeq \Gamma_N \simeq H$;
it was the mechanism that gave rise to the viable parameter regions identified in the earlier study \cite{Canetti:2012vf,Canetti:2012kh}.
This region is substantially modified when taking into account the considerable improvement in the QKEs over the past decade \cite{Biondini:2017rpb,Laine:2022pgk,Eijima:2017anv, Ghiglieri:2017gjz,Ghiglieri:2020ulj,Garbrecht:2018mrp}:
For inverted neutrino mass ordering of neutrino masses (IO) it is entirely removed, for normal ordering (NO) it remains feasible for $4$ GeV $\lesssim M \lesssim 50$ GeV; the upper bound on $M$ is for the first time identified in the present study.
The \emph{ignition mechanism},
a self-amplified growth of $\mu_\alpha$ driven by the  Mikheyev--Smirnov--Wolfenstein (MSW) effect \cite{Wolfenstein:1977ue, Mikheev:1986gs},
is a genuinely new way of generating large $\mu_\alpha$ that we find in this work. It re-opens the HNL mass range below the masses of B-mesons.

\begin{figure}[t]
  \centering
  \includegraphics[width=0.95\columnwidth]{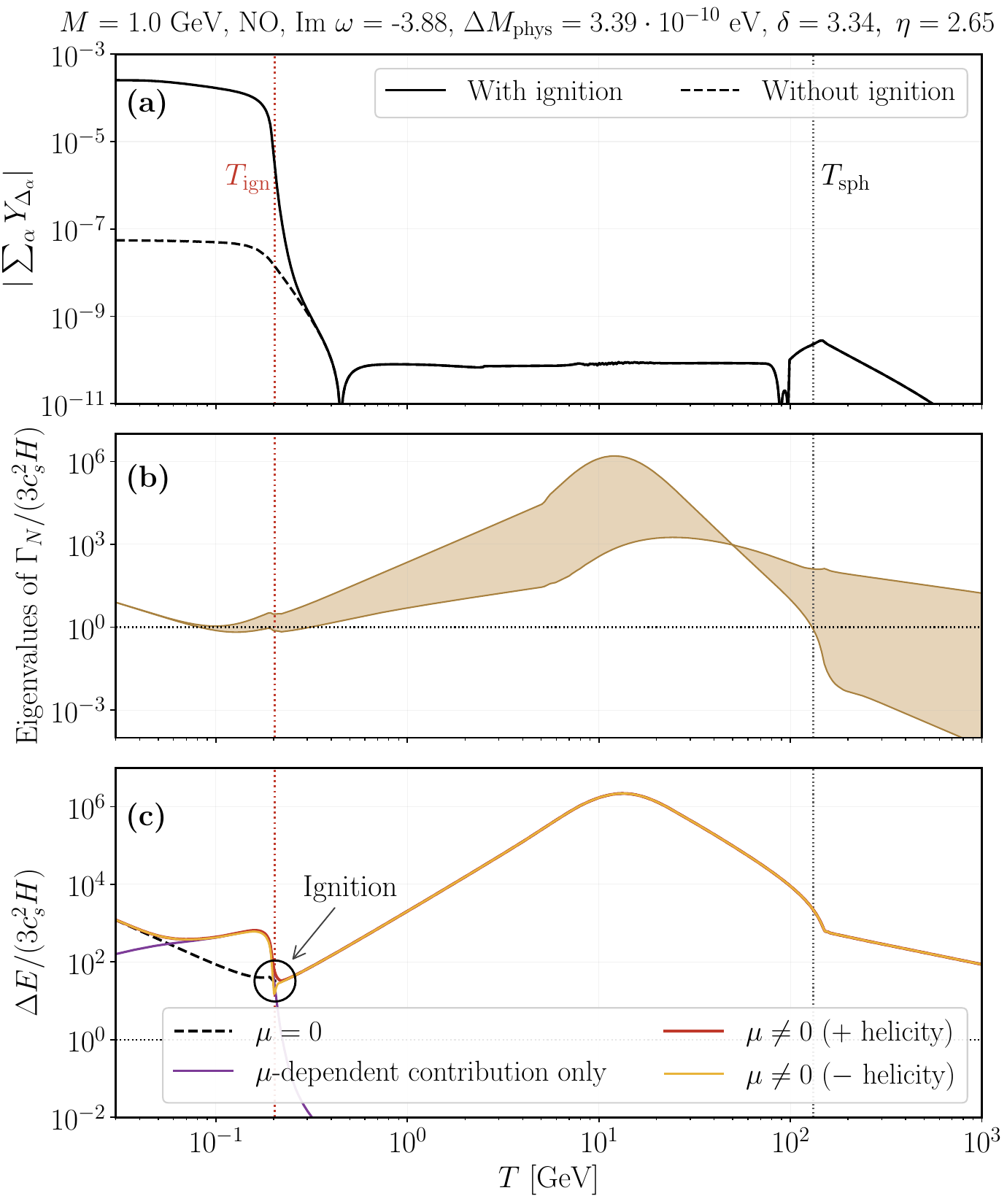}
  \caption{Time evolution for a specific benchmark of a) the absolute value of the total asymmetry $|\sum_\alpha Y_{\Delta_\alpha}|$ charge, b) the 2 eigenvalues of the HNL interaction rates relative to the Hubble rate, and c) the physical in-medium mass splittings determining $\omega_{\rm osc}$ for positive and negative helicity HNLs. 
  In panel a) and c), we compare the evolution of lepton asymmetries and physical mass splittings with (continuous) and without (dashed) including the $\mu_\alpha$-dependent contributions to $H_N$.
 The vertical red and black dotted lines indicate the temperatures $T_{\rm ign}$ and $T_{\rm sph}$. 
 Casas-Ibarra parameters defining the chosen benchmark are specified in the title, using the same convention as in \cite{Klaric:2021cpi}.}
  \label{fig:ignition}
\end{figure}

In vacuum, the HNL masses receive two contributions, one from the Majorana mass matrix in the Lagrangian and one from the Higgs mechanism, the latter is parametrically of the same order $\sim \theta^2 M$ as the light neutrino masses and can enable $\Delta M_{\rm phys}$ below the light neutrino  mass splittings \cite{Canetti:2012kh}. 
In the early universe the $N_i$ dispersion relations are further modified by screening, in particular forward scattering, resulting in an effective Hamiltonian $H_N$ that depends on $T$ and the $\mu_\alpha$. 
The $\mu_\alpha$-dependent piece is analogous to the matter-potential causing the
MSW effect in neutrino kinetic theory \cite{Notzold:1987ik, Sigl:1993ctk}; 
while its importance for DM production has been known for a long time \cite{Shi:1998km, Laine:2008pg}, its impact on the generation of lepton asymmetries $\mu_\alpha$ has not been included in any parameter space scan of the $\nu$MSM.\footnote{The term was acknowledged in the HNL context in \cite{Canetti:2012kh} but not included in the analysis; the kinetic equations of refs.~\cite{Ghiglieri:2017gjz, Ghiglieri:2019kbw, Ghiglieri:2020ulj} contain it as the coefficient functions odd in the chemical potentials. Our results agree with those of \cite{Ghiglieri:2020ulj} at the benchmark point specified there.} 
It is, however, of crucial importance in the regime 
where resonant sterile neutrino DM production is possible, as it turns a plasma with charge asymmetries $\mu_\alpha$ into a birefringent medium for the HNL pair $N_{2,3}$ in which their helicity eigenstates rotate in opposite directions.
Since the two HNL helicities decay preferably into matter and antimatter, this effectively introduces CP-violation in further HNL decays, caused by the matter potential rather than  the phases in their Yukawa couplings.
Moreover, it can enhance the splitting $\Delta E$ between the HNL dispersion relations (given by the eigenvalues of $H_N$) in a regime where the $\mu_\alpha$-independent matter potential is smaller than the vacuum mass splitting, triggering CP-violating HNL oscillations.
Since these asymmetry-generating effects are caused by the $\mu_\alpha$ themselves, 
this leads to a self-amplified growth of the chemical potentials $\mu_\alpha$ once they are large enough to ignite this instability at $T=T_{\rm ign}$,\footnote{ 
The system exhibits non-linear behaviour in spite of the fact that $\mu_\alpha/T\ll 1$ at all times because the asymmetry production is highly sensitive to the HNL dispersion relations.}
amplifying the lepton asymmetries by almost three orders of magnitude for the benchmark in Fig.~\ref{fig:ignition}.
The maximal asymmetries we found are $|\mu_\alpha| \simeq 10^{-3}$.

The asymmetries $\mu_\alpha$ then
create a matter potential for active neutrinos that leads to a level-crossing with $N_1$, causing their resonant conversion into DM at $T\simeq0.4$--$0.2$ GeV through the Shi-Fuller mechanism \cite{Shi:1998km, Laine:2008pg}.  The  $\mu_\alpha$ affect the two $N_1$-helicities with opposite signs, so the level-crossing condition is met for one helicity only, and the flavour carrying the largest $|\mu_\alpha|$ dominates.
Asymmetry generation and DM production thus operate in the same epoch, and the former supplies exactly what the latter requires.  Conversion, in turn, consumes the charge that fuels it.

\begin{figure*}[t]
  \centering
  \includegraphics[width=0.48\textwidth]{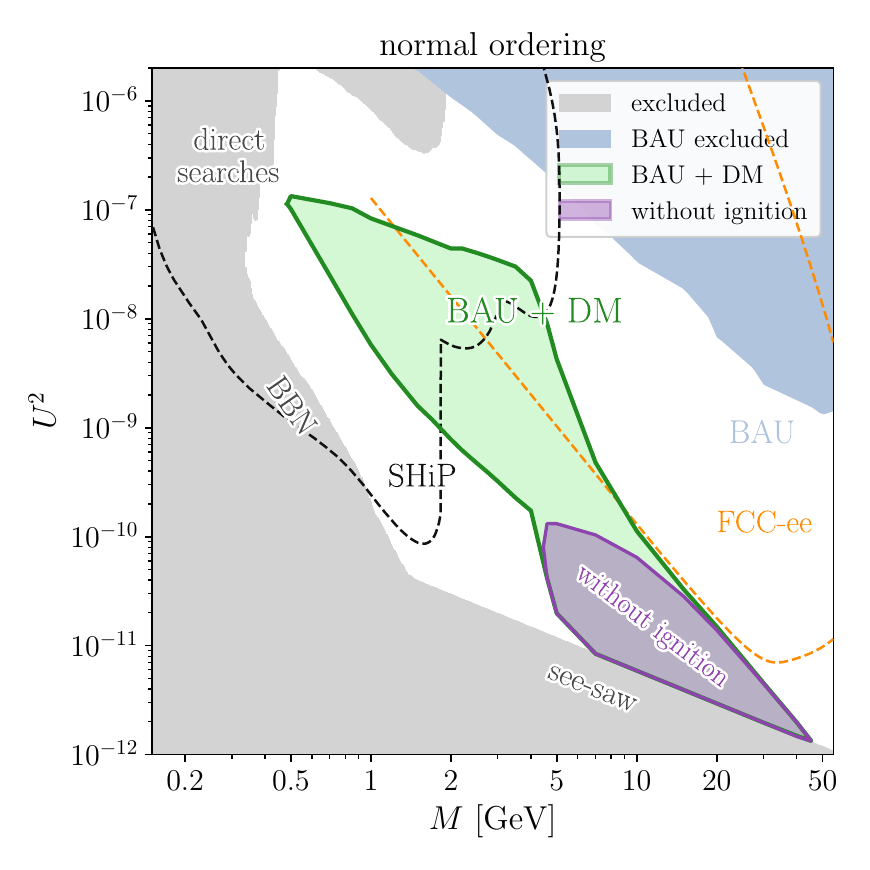}\hfill
  \includegraphics[width=0.48\textwidth]{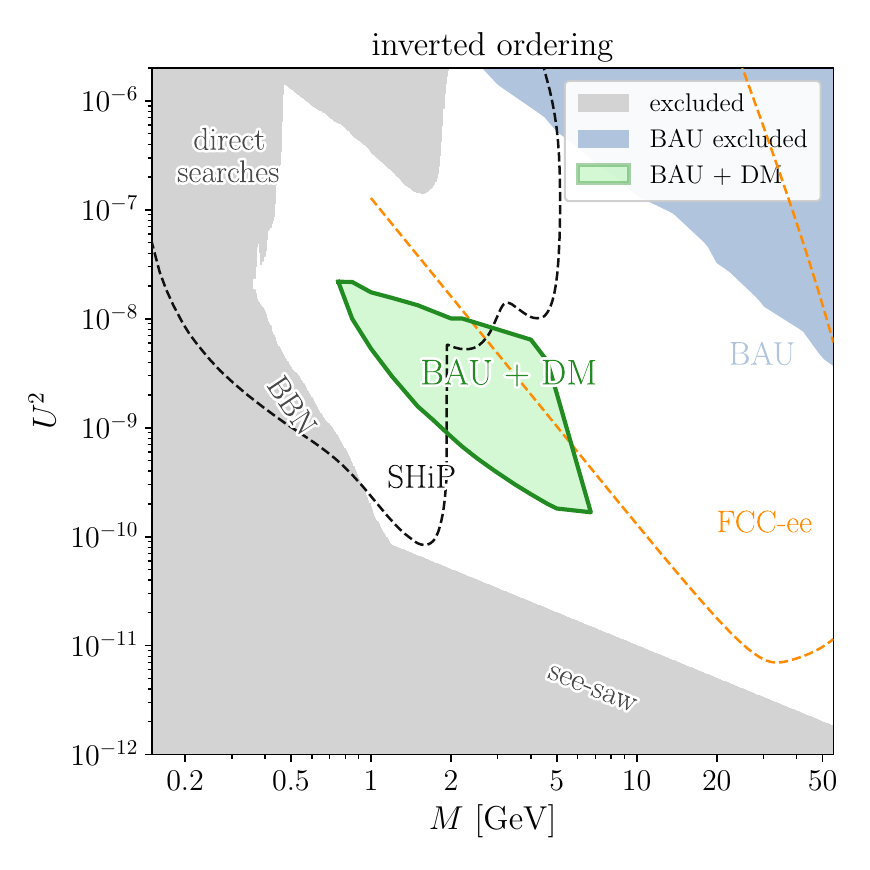}
  \caption{\emph{Green}: Range of mass and mixing where the $\nu$MSM can simultaneously explain the abundances of dark and ordinary matter for NO (left) and IO (right).
  \emph{Violet}: Corresponding region when the ignition mechanism was omitted.
  \emph{Blue}: BAU cannot be explained within the  $\nu$MSM \cite{Klaric:2020phc,Klaric:2021cpi}.
  \emph{Gray}: Excluded by the consistent combination of existing constraints  \cite{Bondarenko:2021cpc}, with dominant bounds in each region indicated in words.
  The lines indicate  projected sensitivities of SHiP \cite{deBlas:2025gyz} (assuming a $\mu$-dominated coupling for NH and an $e$-dominated coupling for IH) and FCC-ee \cite{Drewes:2025ocf}, which could be further improved with additional far detectors \cite{Gninenko:2012anz,Chrzaszcz:2020emg}.
  }
  \label{fig:mixing}
\end{figure*}

\emph{The results.} We solved the QKEs for the HNL density matrices $\rho_N$ and $\bar{\rho}_N$ and the lepton asymmetries from above the electroweak crossover to below the QCD scale, in order to identity the parameter region where the $\nu$MSM can simultaneously explain the observed DM and BAU.
Consistency with neutrino oscillation data is guaranteed by using the Casas-Ibarra parametrisation for the Yukawa matrices $F$ \cite{Casas:2001sr};
we fix all known parameters to their best-fit value \cite{Esteban:2020cvm}
and vary the unknowns freely.
A point passes if the final $Y_B$ matches the observed $\Yobs=8.7\times10^{-11}$ \cite{Aghanim:2018eyx} within a factor of two and the $N_1$ abundance 
reaches the observed dark-matter density 
\cite{Aghanim:2018eyx} at the benchmark $M_1=7$ keV.

 Fig.~\ref{fig:mixing} shows the result in terms of the $N_{2,3}$-pair mass $M$ and its total mixing with the active neutrinos $U^2$. 
For NO there are two separate parameter regions in which the two mechanisms 
-- \emph{resonant decay mechanism} and 
the \emph{ignition mechanism} --
are at work; superimposing them yields the allowed green islands in Fig.~\ref{fig:mixing}.
Roughly speaking, the consistent inclusion of all effects in the QKE pushes the regime where the resonant decay mechanism can create the conditions for resonant DM production to values of $M$ above the B-hadron masses (violet region). The requirement that the HNLs do not decay too early imposes upper limits on $M$ and $U^2$, together with the \emph{seesaw floor} $U^2 > \sum_i m_i/M$ from neutrino oscillation data \cite{Alekhin:2015byh,Drewes:2019mhg,SHiP:2025ows}  they impose an upper bound $M<50$ GeV.
The ignition mechanism re-opens the parameter region $M<5$ GeV, down to $M$ slightly above the kaon mass.
For IO, the resonant decay mechanism is not able to generate the conditions for DM production; as a result, the viable parameter region is restricted to $0.8$~GeV $< M < 7$ GeV. 
We restrict the scan to values $M\gtrsim0.5$~GeV due to   experimental bounds, and uncertainties in the rates .

The parameter region where the $\nu$MSM can simultaneously explain the origins of both dark and baryonic matter entirely falls into the mass range accessible in accelerator-based experiments. 
For $M<2$ GeV (where HNLs are primarily produced in charmed meson decays) SHiP \cite{deBlas:2025gyz} will completely cover all the admitted region. 
For for $2<M<5$ GeV (between the charm and beauty thresholds) a significant part of it will be explored.
 For larger masses the HL-LHC takes over \cite{Izaguirre:2015pga,Drewes:2019fou};\footnote{The sensitivity of the LHC detectors to HNLs produced in meson decays is hampered by their comparably small transverse momenta $p_T$, but could be considerably improved by lowering the $p_T$ cuts \cite{Drewes:2019vjy,CMS:2024zhe}. Additional far detectors can further increase the reach to smaller $U^2$\cite{Agrawal:2021dbo,Abdullahi:2022jlv,Antel:2023hkf,PBC:2025sny}.} 
displaced searches at Tera-Z factories \cite{Blondel:2014bra,Blondel:2022qqo} can extend this reach close to the see-saw floor \cite{Drewes:2025ocf}.

\begin{figure}[t]
  \centering
  \includegraphics[width=0.95\columnwidth]{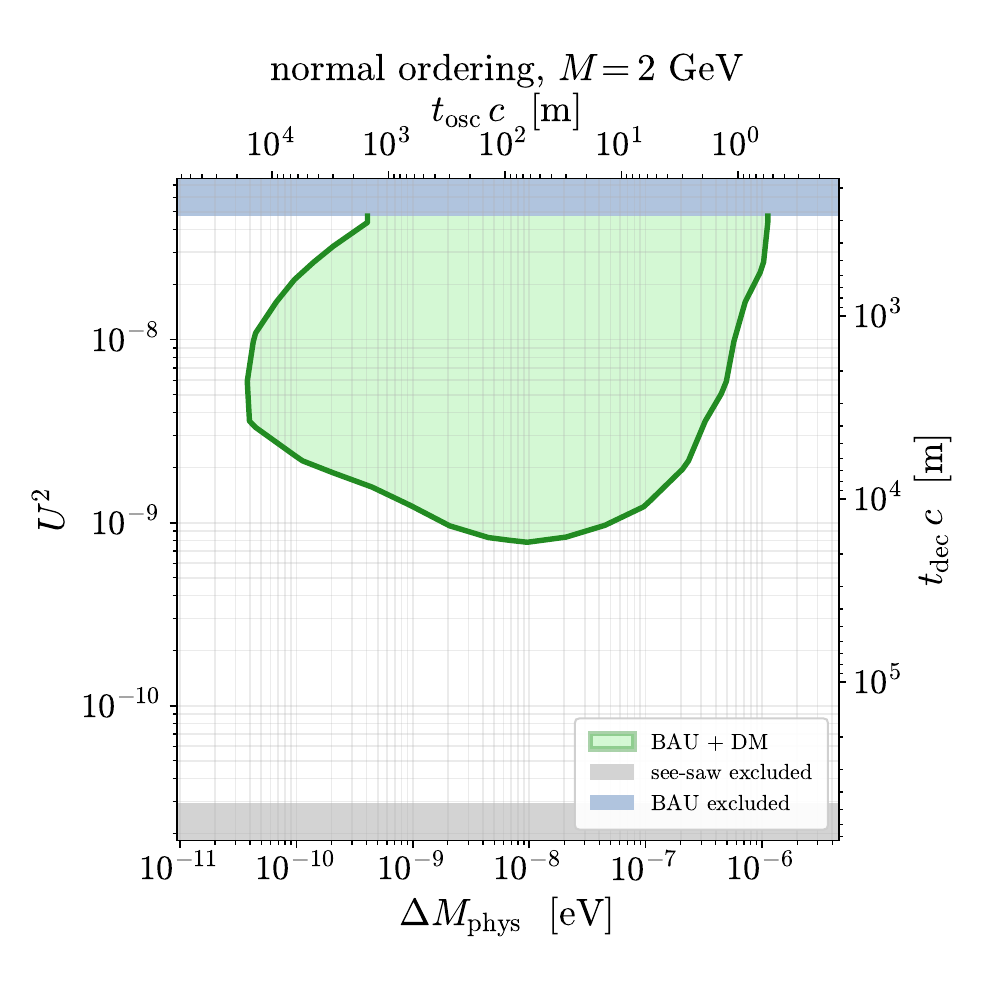}
  \caption{
  Range of allowed mixing angles $U^2$ and physical mass splitting $\Delta M_{\rm phys}$ 
  between $N_2$ and $N_3$ for a benchmark mass of $M=2$ GeV consistent with baryogenesis the generation of the observed DM abundance (green) or baryogenesis alone (white) with NO.
  $t_{\rm osc} c$ indicates the length
  of coherent $N_{2,3}$ oscillations in their rest frame observable at SHiP \cite{Tastet:2019nqj}, IO yields km-length $c t_{\rm osc}$.  }
  \label{fig:splitting}
\end{figure}

Inside the discovery regions the number of events rapidly increases with $U^2$ \cite{Drewes:2019vjy,Bondarenko:2019yob,Drewes:2022rsk}, implying that their properties can be studied in detail. The requirement for successful baryogenesis already implies testable predictions for $N_{2,3}$ decays at colliders (lifetimes, branching ratios, and LNV) \cite{Hernandez:2016kel,Drewes:2016jae,Antusch:2017pkq} as well as neutrinoless double $\beta$-decay \cite{Drewes:2016lqo,Hernandez:2016kel,deVries:2024rfh}. The prediction for $\Delta M_{\rm phys}$ in Fig.~\ref{fig:splitting} opens an additional window to test the mechanism of DM production.  
The pair acts as one quasi-Dirac particle whose decays can exhibit HNL oscillations over macroscopic distances, modulating the 
fraction of LNV decays
that can be observed at SHiP \cite{Tastet:2019nqj} or FCC-ee \cite{Antusch:2024otj}.\footnote{Here we neglect possible decoherence from the wave packets \cite{Antusch:2023nqd}.} 
For NO the oscillation length can fall within the SHiP decay volume, see Fig.~\ref{fig:splitting}; for IO  the pair would appear as a single Dirac-like particle, deviating from the generic expectation in this mass region  \cite{Drewes:2019byd}.  The degree of degeneracy is thus a testable prediction of the DM production mechanism in the $\nu$MSM.

\emph{Conclusions and outlook.}--
The $\nu$MSM has been proposed as an all-in-one solution that can simultaneously explain neutrino masses along with the dark and baryonic matter in the universe by a minimal extension of the SM with three right-handed neutrinos.
This minimality makes the model highly predictive, and efforts to pin down the HNL properties that can be tested in experiments stretch over two decades. 
While constraints from baryogenesis are now well-established (see \cite{Hernandez:2016kel,Drewes:2016jae,Antusch:2017pkq,deVries:2024rfh,Drewes:2016jae} and references therein), no similar predictions exist for the mechanism of DM production -- the validity of the only systematic parameter scan in the literature \cite{Canetti:2012vf,Canetti:2012kh} has been but into question by major developments in the understanding of the QKE governing the HNL dynamics in the early universe. With the present Letter we close this gap and provide the first scan of the $\nu$MSM parameter space using state-of-the-art QKEs that consistently follows the HNL evolution throughout cosmic history from baryogenesis to DM production.
Here we focus on reporting our main results, we will present further details on our analysis and predictions beyond our DM benchmark in an accompanying article.

Our results are surprising and provide important input for HNL searches at accelerators.
Assuming no other new physics beyond the SM, standard General Relativity and a standard cosmic history with vanishing initial HNL abundances -- as e.g.~predicted by Higgs inflation -- the $\nu$MSM makes accurate predictions for the HNL properties based on the requirement to simultaneously explain the dark and baryonic matter abundances along with neutrino oscillations. 
Most importantly, while the parameter space for baryogenesis extends to almost arbitrarily large massses $N$ \cite{Klaric:2020phc,Klaric:2021cpi,Klaric:2021cpi}, the requirement to create the conditions for DM production restricts $M$ to values below the electroweak scale, making it kinematically accessible to beam dump and collider experiments.
Moreover, the HNL mass splitting is predicted to lie in a range where coherent HNL oscillations could be observed could be observed at SHiP or FCC-ee, adding a testable prediction that is specific to the mechanism of DM production to those for other HNL properties from baryogenesis.

On the dark-matter side, the model predicts a specific spectrum of DM sterile neutrinos. The produced $N_1$ population is generally colder than thermal, so structure-formation limits must be evaluated on the computed spectra \cite{Boyarsky:2018tvu, Schneider:2016uqi}.
Finally, the large lepton asymmetries  $|\mu_\alpha|10^{-3}$ may impact other cosmological observables such as BBN \cite{Domcke:2025jiy} or the rate of primordial black hole formation \cite{Bodeker:2020stj}.

\begin{acknowledgments}

The authors would like to thank Shintaro Eijima, Jacopo Ghiglieri, Mikko Laine and Maksym Ovchynnikov for helpful discussions. 
The work of MaD has been partially funded by the Deutsche Forschungsgemeinschaft (DFG, German Research Foundation) - SFB 1258 - 283604770.  The work of Y.G. has been supported by the World Premier International Research Center Initiative (WPI), MEXT, Japan (Kavli IPMU).
The work of J.K. has been supported by the HRZZ grant UIP-2025-02-3909 and the Ministry of Science, Education and
Youth (MZOM) Multilateral Scientific and Technological Cooperation in the Danube Region project 910-06/25-01/00041.
The work of I.T. has been supported by a research grant (VIL57416) from VILLUM FONDEN.
Computational resources have been provided by the supercomputing facilities of the Université catholique de Louvain (CISM/UCL) and the Consortium des Équipements de Calcul Intensif en Fédération Wallonie Bruxelles (CÉCI) funded by the Fond de la Recherche Scientifique de Belgique (F.R.S.-FNRS) under convention 2.5020.11 and by the Walloon Region,  and by the NVIDIA Academic Grant Program. Y.G. also acknowledges the use of the \textit{idark} cluster at IPMU.
J.K. acknowledges the use of the Supek supercomputer at the University of Zagreb computing centre SRCE.
Artificial Intelligence (AI) systems (Claude Fable~5 and~5.1, Claude Opus family, GPT-5.6 Sol, GPT-6 Astra) supervised by the authors were used for coding, performing parameter space scans and discussing physics results, including comparing them to the literature. The results of the AI generated code were compared against a human-written code. 

 \end{acknowledgments}

\bibliography{references}

\clearpage
\onecolumngrid

\section*{Supplemental Material on Quantum kinetic equations}\label{sec:QKE}

The dynamical quantities relevant in the present context comprise the density matrices $\rho_N$ and $\rho_{\bar{N}}$ for the two helicity states of the HNL pair $N_{2,3}$,
the abundances $f_1^\pm$ of the to $N_1$ helicities, and a set of chemical potentials
$\mu\equiv(\mu_e,\mu_\mu,\mu_\tau,\mu_Q,\mu_Z)$ for lepton flavour, electric charge, and effective $Z$-charge.
The QKE's describing their time evolution \cite{Ghiglieri:2019kbw,Bodeker:2019rvr}\footnote{See e.g.~\cite{Garbrecht:2018mrp,Klaric:2021cpi} for reviews of earlier developments.} formally resemble the density matrix equations known from neutrino physics \cite{Sigl:1993ctk}. 
Here we display their structure, introduce our notation and sketch the underlying physics; a more detailed derivation will be provided in a follow-up work.  
Per comoving momentum mode $k$, the positive-helicity density matrix $\rho_N$ obeys
\begin{equation}
i\,\frac{{\rm d}\rho_N}{{\rm d}t}
 = \big[H_N,\rho_N\big]
 -\frac{i}{2}\big\{\Gamma_N,\delta\rho_N\big\}
 -\frac{i}{2}\sum_\alpha\tilde\Gamma_\alpha\, 2\frac{\mu_\alpha}{T}f_N(1-f_N)
\label{eq:rhoN}
\end{equation}
where $\delta\rho_N\equiv\rho_N-f_N\,\mathds{1}$ is the departure from equilibrium ($f_N$ the Fermi--Dirac distribution). 
The commutator drives coherent $N_2$--$N_3$ oscillations, 
the remaining terms in \eqref{eq:rhoN} comprise the collision integral. Amongst them, the anticommutator describes the damping of deviations from equilibrium at the rate $\Gamma_N$, and the last term is the back-reaction of the flavoured asymmetries $\mu_\alpha$. 

The effective Hamiltonian can be parameterised as 
\begin{equation}
H_N(\mu)=H_0
 +\sum_\alpha\big[\mathfrak{h}_{+ \alpha}(\mu)\,Y^\alpha_+ + \mathfrak{h}_{- \alpha}(-\mu)\,Y^\alpha_-\big],
\label{eq:HN}
\end{equation}
with $H_0={\rm diag}[(\textbf{k}^2+M_2^2)^{1/2},(\textbf{k}^2+M_3^2)^{1/2}]$ the free Hamiltonian
and $\mathfrak{h}_{\pm \alpha}$ coefficients 
$(Y^\alpha_+)_{IJ}=F^*_{\alpha I}F_{\alpha J}$ 
and $Y^\alpha_-=(Y^\alpha_+)^*$ flavour matrices,
and $\mathfrak{h}_{\pm \alpha}$ coefficients that define the in-medium dispersion relations for the various components in flavour space.\footnote{In the massless limit it reduces to the Wolfenstein-type matter potential of active-neutrino kinetics \cite{Notzold:1987ik, Sigl:1993ctk}.}
While in general all coefficients in the QKE depend on momentum, temperature and the chemical potentials, we indicate the dependence on $\mu$ explicitly in \eqref{eq:HN} because of its significance for the ignition mechanism.
The rate matrices $\Gamma_N$ and $\tilde\Gamma_\alpha$ comprising the collision term are built the same way from absorptive coefficients $\gamma_{\pm \alpha}$.\footnote{Note that we have expanded the collision term to linear order in $\mu/T$ to obtain a manifestly closed set of equations in \eqref{eq:rhoN}, so that the $\gamma_{\pm \alpha}$ are (in contrast to the $\mathfrak{h}_{\pm \alpha}$) independent of $\mu$.}
These thermal coefficients were primarily computed in Fermi theory, assuming free quarks.
This description breaks down for masses $M\lesssim 500$ MeV, as the HNL decay into mesons below the QCD crossover, and for $M\gtrsim 50$ GeV,
where Fermi theory is no longer valid. This introduces a moderate uncertainty on both edges of the mass range.
The negative-helicity matrix $\rho_{\bar N}$ obeys the same equation with all charges reversed, the coefficients conjugated, $H_N(\mu)\to H_N(-\mu)^*$. 
In practice, we solve the QKEs for the CP-even and odd deviations from equilibrium instead of $\rho_N$ and $\bar{\rho}_N$.

The various chemical potentials $\mu$ are connected via a susceptibility matrix \cite{Bodeker:2015zda,Garbrecht:2019zaa}, so that only three flavoured charges need to be tracked dynamically,
\begin{equation}
\frac{{\rm d}Y_{\Delta_\alpha}}{{\rm d}t}
 = \frac{1}{s}\!\int_{\mathbf k}\!\Big\{
 -\Gamma_{\nu_\alpha}g_\alpha
 +{\rm Tr}\big[\tilde\Gamma_\alpha(-\mu)^*\delta\rho_{\bar N}
              -\tilde\Gamma_\alpha(\mu)\,\delta\rho_N\big]\Big\},
\label{eq:charges}
\end{equation}
where $Y_{\ldots}\equiv n_{\ldots}/s$ with $s$ the entropy density and $n_{\ldots}$ the corresponding number density.  
The baryon charge follows the sphaleron equation, ${\rm d}Y_B/{\rm d}t=-\Gamma_B(T)[Y_B+\chi(T)\sum_\alpha Y_{\Delta_\alpha}]$ \cite{Eijima:2017cxr}, whose rate shuts off dynamically at $T_{\rm sph}$, 
and the $\mu_\alpha$ follow from the charges $Y_{\Delta_\alpha}$ through the susceptibility matrix $\chi$ of the SM plasma.

Finally, the asymmetries convert into dark matter through \cite{Shi:1998km, Laine:2008pg, Ghiglieri:2020ulj}
\begin{equation}
\frac{{\rm d}f_1^{(\pm)}(k)}{{\rm d}t}
 = \sum_\alpha \Gamma^{(\pm)}_{1 \alpha}(T,k;\mu)\,
   \big[n_F((\textbf{k}^2+M_1^2)^{1/2}\pm\mu_\alpha)-f_1^{(\pm)}\big],
\label{eq:N1}
\end{equation}
where the rate $\Gamma^{(\pm)}_{1 \alpha}\propto\sin^2(2\theta_1)$ is resonantly enhanced when the asymmetry-induced active neutrino matter potential causes a level-crossing with $N_1$.

We solve the system \eqref{eq:rhoN}-\eqref{eq:N1}  with full momentum dependence in \eqref{eq:N1}. When searching for viable parameter regions, we solve momentum-averaged versions of \eqref{eq:rhoN} and the corresponding equation for $\rho_{\bar{N}}$, and we neglect the depletion of the $\mu_\alpha$ due to $N_1$-production. We then cross-check a posteriori that these approximations were justified on the boundaries of the islands in Fig.~\ref{fig:mixing}.
We include the expansion of the universe, dynamical sphaleron freeze-out, the entropy release of the QCD epoch and of the HNL decays, hadronic contributions to the susceptibilities and potentials, and the back reaction of the asymmetries and of the DM transfer on the heavy sector.  Each previous analysis \cite{Canetti:2012vf, Eijima:2020shs, Klaric:2021cpi, Ghiglieri:2020ulj} omitted at least one of these. A decisive new element in the present work lies in the consistent inclusion of the charge-odd part in \eqref{eq:HN}.

\end{document}